\documentclass[12pt]{article}

\usepackage{amsmath,amsfonts,amssymb,latexsym}

\newtheorem{theorem}{Theorem}[section]

\newenvironment{proof}[1][Proof]{\textsc{#1.} }{\ \rule{0.5em}{0.5em}}
\numberwithin{equation}{section}
\def\be{\begin{equation}}
\def\ee{\end{equation}}
\def\bq{\begin{eqnarray}}
\def\eq{\end{eqnarray}}
\def\beq{\begin{eqnarray*}}
\def\eeq{\end{eqnarray*}}

\begin{document}

\begin{titlepage}
\begin{flushright}
%add
\end{flushright}

\vspace{1cm}

\begin{center}
{\huge Cosmological Singularities and Bel-Robinson Energy}

\vspace{1cm}

{\large Spiros Cotsakis$\dagger$ and Ifigeneia
Klaoudatou$\ddagger$}\\

\vspace{0.5cm}

{\normalsize {\em Research Group of Geometry, Dynamical Systems
and Cosmology}}\\ {\normalsize {\em Department of Information and
Communication Systems Engineering}}\\ {\normalsize {\em University
of the Aegean}}\\ {\normalsize {\em Karlovassi 83 200, Samos,
Greece}}\\ {\normalsize {\em E-mails:} $\dagger$
\texttt{skot@aegean.gr}, \texttt{$\ddagger$ iklaoud@aegean.gr}}
\end{center}

\vspace{0.7cm}
\begin{abstract}
\noindent We consider the problem of describing the asymptotic
behaviour of \textsc{FRW} universes near their spacetime
singularities in general relativity. We find that the Bel-Robinson
energy of these universes in conjunction with the Hubble expansion
rate and the scale factor proves to be an appropriate measure
leading to a complete classification of the possible singularities.
We show how our scheme covers all known cases of cosmological
asymptotics possible in these universes and also predicts new and
distinct types of singularities. We further prove that various
asymptotic forms met in flat cosmologies continue to hold true in
their curved counterparts. These include phantom universes with
their recently discovered big rips, sudden singularities as well as
others belonging to graduated inflationary models.
\end{abstract}

%\vspace{0.3cm}
%\begin{center}
%{\line(5,0){280}}
%\end{center}

\end{titlepage}
%\tableofcontents
%\newpage
\section{Introduction}
The problem of cosmological singularities is perhaps the single
most important unsolved issue in modern theoretical general
relativity and cosmology. It is usually formulated as being a
two-faceted problem, the first aspect being the issue of finding
conditions under which singularities are expected to develop in a
finite time (existence) during the evolution of a model spacetime,
the second being that of describing their character or nature. In
fact, the other, closely related, issue in general relativity,
that of proving geodesic completeness, spacetimes existing for an
infinite proper time interval, is usually thought of as the
negation of the singularity problem, cf. \cite{cbm03}. The
geodesic completeness problem may also be formulated as  a
two-sided issue, one aspect being, like the singularity problem,
that of finding conditions that guarantee the existence of
complete spacetimes, the other being that of unraveling the nature
of geodesic completeness, that is the nature of behaviour of the
controlling functions, for instance, their decay to infinity.

The issue of proving existence of singular and/or complete
spacetimes is basically a geometric one. Traditionally one is
interested in formulating criteria, of a geometric nature, to test
under what circumstances singularities, in the form of geodesic
incompleteness, will be formed during the evolution in general
relativity and in other metric theories of gravity, cf.
\cite{he73}. These, apart from the causality ones like global
hyperbolicity, are usually translated into sufficient conditions
to be satisfied by the matter fields present and are very
plausible. Such conditions guarantee that in a finite time in the
future or past of the spacetime, some quantity, usually describing
the convergence of a bundle of geodesics, blows up thus
designating the past or future breakdown of the bundle. On the
other hand, one can  formulate equally plausible and generic
geometric criteria for the long time existence of geodesically
complete, generic spacetimes in general relativity and other
theories of gravity, cf. \cite{chc02}. Such sufficient conditions
assume a globally hyperbolic spacetime in the so-called generic
sliced form (cf. \cite{c04}), require the space gradient of the
lapse function as well as the extrinsic curvature not to grow
without bound, and are also very plausible. They provide us with
assumptions under which generic spacetimes will exist forever. All
known global-in-time spacetimes satisfy these sufficient
conditions and are therefore future geodesically complete.

Of course as  the singularity theorems do not detail the nature of
the behaviour of the controlling functions on approach to the
singularity, similarly the completeness theorems do not provide
conditions for, say, the decay of such functions to infinity. A
lesson to be drawn from this state of affairs is that in general
relativity (and in other metric gravity theories), singular and
complete spacetimes are equally generic in a sense. The real
question is what do we mean when we say that a relativistic model
develops a spacetime singularity during its evolution. In other
words, what are the possible spacetime singularities which are
allowed in gravity theories? These may be singularities in the
form of geodesic incompleteness developing in the course of the
evolution but can be also others which are more subtle and which
are also dynamical and present themselves, for instance, in some
higher derivatives of the metric functions spoiling the smoothness
of global solutions to the field equations.

Obviously a recognition and complete analysis of such a program
cannot be accomplished in the short run and requires complete
examination of a number of different factors controlling the
resulting behaviour. For instance, one needs to have control of
each possible behaviour of the different families of relativistic
geometry coupled to matter fields in general relativity and other
theories of gravity to charter the possible singularity formation.
In this sense, unraveling the nature and kinds of possible
singularities in the simplest kinds of geometry becomes equally
important as examining this problem in the most general solutions
to the Einstein equations. In fact, an examination  of the
literature reveals that the more general the spacetime geometry
considered (and thus the more complex the system of equations to
be examined) the simplest is the type of singularities allowed to
be examined. By starting with a simple cosmological spacetime we
allow all possible types of singularities to come to surface and
be analyzed.

In \cite{ck05} we derived necessary conditions for the existence
of finite time singularities in globally and regularly hyperbolic
isotropic universes, and provided first evidence for their nature
based entirely on the behaviour of the Hubble parameter (extrinsic
curvature). This analysis exploited directly the evidence provided
by the completeness theorems proved in \cite{chc02}. The main
result proved in \cite{ck05} may be summarized as follows:
\begin{theorem}\label{1}
Necessary conditions for the existence of finite time singularities in
globally hyperbolic, regularly hyperbolic \textsc{FRW} universes are:
\begin{description}
\item[$S_{1}$] For each finite $t$, H is non-integrable on $[t_1,t]$,
or
\item[$S_{2}$] H blows up in a finite future time, or
\item[$S_{3}$] H is defined and integrable for only a finite
proper time interval.
\end{description}
\end{theorem}

Condition $S_{1}$ may describe different types of singularities.
For instance, it describes  a big bang type of singularity, when
$H$ blows up at $t_{1}$ since then it is not integrable on any
interval of the form $[t_{1},t]$,  $t>t_{1}$ (regular
hyperbolicity is violated in this case, but the scale factor is
bounded from above). However, under  $S_{1}$ we can have other types of
singularities: Since $H(\tau)$ is integrable on an interval
$[t_1,t]$, if $H(\tau)$ is defined on $[t_1,t]$,  continuous on
$(t_1,t)$ and the limits $\lim_{\tau\rightarrow t_1^+}H(\tau)$ and
$\lim_{\tau\rightarrow t^-}H(\tau)$ exist, the violation of
\emph{any} of these conditions leads to a singularity that is not
of the big bang type discussed previously.

Condition $S_{2}$ describes
a future blow up singularity and condition $S_{3}$ may lead to a
sudden singularity (where $H$ remains finite), but for this to be
a genuine type of singularity, in the sense of geodesic
incompleteness, one needs to demonstrate that the metric is
non-extendible to a larger interval.

Note that these three conditions are not overlapping, for example
$S_{1}$ is not implied by $S_{2}$ for if $H$ blows up at some
finite time $t_s$ after $t_1$, then it may still be integrable on
$[t_1,t]$, $t_1<t<t_s$. As discussed in \cite{ck05}, Theorem
(\ref{1}) describes possible time singularities that are met in
$\textsc{FRW}$ universes having a Hubble parameter that behaves
like $S_{1}$ or $S_{2}$ or $S_{3}$. Although such a classification
is a first step to clearly distinguish between the various types
of singularities that can occur in such universes, it does not
bring out some of the essential features of the dynamics that
differ from singularity to singularity. For instance, condition
$S_{2}$ includes both a collapse singularity, where the scale
factor $a\rightarrow 0$ as $t\rightarrow t_{s}$, \emph{and} a blow
up singularity where $a\rightarrow \infty$ as $t\rightarrow
t_{s}$. Such a degeneration is unwanted in any classification  of
the singularities that can occur in the model universes in
question.

It is therefore necessary to extend and refine this classification
by considering also the behavior of the scale factor. That it is
only necessary to include this behaviour, is also seen most
clearly by noticing that only in this way one may consistently
distinguish between initial and final singularities, or between
past and future ones, as it has been repeatedly emphasized by
Penrose  over the years in his Weyl curvature hypothesis, cf.
\cite{pen04}. Another aspect of the problem that must be taken
into account is the relative behaviour of the various  matter
components as we approach the time singularities. We believe that
a clear picture of the behaviour of the matter fields on approach
to a singularity will contribute to an understanding of its
nature.

In this paper we present a new scheme that describes in a
consistent way the possible spacetime singularities which can form
during the evolution of the {\sc FRW} cosmological models in
general relativity. Section 2 introduces the classification and
shows that by extending our previous results and taking into
account the effects induced by the Bel-Robinson energy, we can
arrive at the possible forms of singular behaviour that the
universes in question can exhibit. Section 3 proves that various
singularity types predicted by our scheme have a general
significance in the isotropic category and also describes how some
singularity types known to occur only in flat isotropic universes
do in fact appear in many curved spacetimes. In the last Section
we discuss various  aspects of this work including the issue of
how to describe known types of singularities with our technique.

\section{Classification}

In this Section we present a classification of possible
singularities which occur in isotropic model universes. Our
classification is based on the introduction of an invariant
geometric quantity, the Bel-Robinson energy (cf. \cite{ycb02} and
Refs. therein), which takes into account precisely those features
of the problem, related to the matter contribution, in which
models still differ near the time singularity while having similar
behaviours of $a$ and $H$. In this way, we arrive at a complete
classification of the possible cosmological singularities in the
isotropic case.

The Bel-Robinson energy is a kind of energy of the gravitational
field \emph{projected} in a sense to a slice in spacetime. It is
used in \cite{ck93}, \cite{chm} to prove global existence results
in the case of an asymptotically flat and cosmological spacetimes
respectively, and is defined as follows. Consider a sliced
spacetime with metric
\begin{equation}
^{(n+1)}g\equiv -N^{2}(\theta ^{0})^{2}+g_{ij}\;\theta ^{i}\theta ^{j},\quad
\theta ^{0}=dt,\quad \theta ^{i}\equiv dx^{i}+\beta ^{i}dt,  \label{2.1}
\end{equation}
where $N=N(t,x^{i})$ is the lapse function and $\beta^{i}(t,x^{j})$ is the
shift function, and the $2$-covariant spatial electric
and magnetic tensors
\beq
E_{ij}&=&R^{0}_{i0j},\\
D_{ij}&=&\frac{1}{4}\eta_{ihk}\eta_{jlm}R^{hklm},\\
H_{ij}&=&\frac{1}{2}N^{-1}\eta_{ihk}R^{hk}_{0j},\\
B_{ji}&=&\frac{1}{2}N^{-1}\eta_{ihk}R^{hk}_{0j},
\eeq
where
$\eta_{ijk}$ is the volume element of the space metric $\bar g$. These
four time-dependent space tensors comprise what is called a
\emph{Bianchi field}, $(E,H,D,B)$, a very important frame field
used to prove global in time results, cf. \cite{ycb02}.

The \emph{Bel-Robinson energy at time $t$} is given by \be
\mathcal{B}(t)=\frac{1}{2}\int_{\mathcal{M}_t}\left(|E|^{2}+|D|^{2}+
|B|^{2}+|H|^{2}\right)d\mu_{\bar{g}_t}, \ee where by
$|X|^{2}=g^{ij}g^{kl}X_{ik}X_{jl}$ we denote the spatial norm of
the $2$-covariant tensor $X$. In the following, we exclusively use
an $\textsc{FRW}$ universe filled with various forms of matter
with metric given by
\be ds^2=-dt^2+a^2(t )d\sigma ^2,
\ee
where
$d\sigma ^2$ denotes the usual time-independent metric on the
3-slices of constant curvature $k$. For this spacetime, we find that the norms
of the magnetic parts, $|H|, |B|$, are identically zero while $|E|$ and
$|D|$, the norms of the electric parts, reduce to the forms
\be
|E|^{2}=3\left(\ddot{a}/{a}\right)^{2} \quad\textrm{and}\quad
|D|^{2}=3\left(\left(\dot{a}/{a}\right)^{2}+k/{a^{2}}\right)^{2}.
\ee Therefore the Bel-Robinson energy becomes \be
\mathcal{B}(t)=\frac{C}{2}\left(|E|^{2}+|D|^{2}\right), \ee where
$C$ is the constant volume of (or \emph{in} in the case of a
non-compact space) the 3-dimensional slice at time $t$.

It is not difficult to show that a closed, \textsc{RW} universe
such that $|D|$ is bounded above, $H$ must be bounded above and
the scale factor bounded below. Therefore $H$ must be integrable
and the spatial metric bounded below, that is such a universe is
regularly hyperbolic. This in turn means that all hypotheses of
the completeness theorem proved in \cite{chc02} are satisfied and
therefore such a universe is $g-$complete. It is also
straightforward to see that the null energy condition is
equivalent to the inequality $|E|\leq |D|$, hence completeness is
then accompanied with $|E|$ being bounded above. (For a flat
spatial metric we would need to impose the regular hyperbolicity
hypothesis in order to conclude completeness since the latter is
independent from the boundedness of $|D|$.)

We can now proceed to list the possible types of singularities
that are formed in an $\textsc{FRW}$ geometry during its cosmic
evolution and enumerate the possible types that result from the
different combinations of the three main functions in the problem,
namely, the scale factor $a$, the Hubble expansion rate $H$ and
the Bel Robinson energy $\mathcal{B}.$ These types will by
necessity entail a possible blow up in the functions $|E|$, $|D|$.
If we suppose that the model has a finite time singularity at
$t=t_s$, then the possible behaviours of the functions in the
triplet $\left(H,a,(|E|,|D|)\right)$ in accordance with Theorem
(\ref{1}) are as follows:
\begin{description}
\item [$S_{1}$] $H$ non-integrable on $[t_{1},t]$ for every $t>t_{1}$

\item [$S_{2}$] $H\rightarrow\infty$  at $t_{s}>t_{1}$

\item [$S_{3}$] $H$ otherwise pathological
\end{description}

\begin{description}
\item [$N_{1}$] $a\rightarrow 0$

\item [$N_{2}$] $a\rightarrow a_{s}\neq 0$

\item [$N_{3}$] $a\rightarrow \infty$
\end{description}

\begin{description}
\item [$B_{1}$] $|E|\rightarrow\infty,\, |D|\rightarrow \infty$

\item [$B_{2}$] $|E|<\infty,\, |D|\rightarrow \infty $

\item [$B_{3}$] $|E|\rightarrow\infty,\, |D|< \infty $

\item [$B_{4}$] $|E|<\infty,\, |D|< \infty $.
\end{description}
The nature of a prescribed  singularity is thus described completely by
specifying the components in a triplet of the form \[(S_{i},N_{j},B_{l}),\]
with the indices $i,j,l$ taking their respective values as above.

Note that there are a few types that cannot occur. For instance,
we cannot have an $(S_{2},N_{2},B_{3})$ singularity because that
would imply having $a<\infty$ ($N_{2}$) and $H\rightarrow\infty$
$(S_{2})$, while $3\left(
(\dot{a}/{a})^{2}+k/a^{2}\right)^{2}<\infty$ $(B_{3})$, at $t_{s}$
which is impossible since $|D|^{2}\rightarrow\infty$ at $t_{s}$
($k$ arbitrary).

A complete list of impossible singularities is model-dependent and
is generally given by triplets  $(S_{i},N_{j},B_{l})$, where the
indices, in the case of a $k=0,+1$ universe, take the values
$i=1,2$, $j=1,2,3$, $l=3,4$, whereas in the case of a $k=-1$
universe the indices take the values $i=1,2$, $j=2,3$, $l=3,4$
(here by $S_{1}$ we denote for simplicity only the big bang case
in the  $S_{1}$ category). We thus see that some singularities
which are impossible for a flat or a closed universe become
possible for an open universe. Consider for example the triplet
$(S_{2},N_{1},B_{3})$ which means having
 $H\rightarrow\infty$, $a\rightarrow 0$ and
$$|D|^2=3\left(\left(\dot{a}/{a}\right)^{2}+k/{a^{2}}\right)^{2}<\infty$$
at $t_{s}$. This behaviour is valid only for some cases of an open universe.

All other types of finite time singularities can in principle be
formed during the evolution of \textsc{FRW}, matter-filled models,
in general relativity or other metric theories of gravity.

It is interesting to note that all the standard dust or
radiation-filled big bang singularities fall under the
\emph{strongest} singularity type, namely, the type
$(S_{1},N_{1},B_{1})$. For example, in a flat universe filled with
dust, at $t=0$ we have \bq a(t)&\propto& t^{2/3}\rightarrow 0,
\quad (N_{1}),\\ H&\propto& t^{-1}\rightarrow\infty, \quad
(S_{1}),\\ |E|^{2}&=&3/4H^{4}\rightarrow\infty,\quad
|D|^{2}=3H^{4}\rightarrow\infty, \quad (B_{1}). \eq Note that our
scheme is organized in such  a way that the character of the
singularities (i.e., the behaviour of the defining functions)
becomes milder as the indices of $S$, $N$ and $B$ increase. Milder
singularities in isotropic universes are thus expected to occur as
one proceeds down the singularity list.

It is the purpose of this classification to apply both to vacuum as
well as matter dominated models. In fact the Bel-Robinson energy takes
care in a very neat way the matter case. For instance, in fluid-filled
models, the various behaviours of the Bel-Robinson energy
density can be related to four conditions imposed on the density and
pressure of the cosmological fluid:
\begin{description}
\item [$B_{1}$] $\Leftrightarrow$ $\mu\rightarrow \infty$ and $|\mu+3p|\rightarrow\infty$

\item [$B_{2}$] $\Leftrightarrow$ $\mu\rightarrow \infty$ and $|\mu+3p|<\infty$

\item [$B_{3}$] $\Leftrightarrow$ $\mu<\infty$ and $|\mu+3p|\rightarrow\infty$
$\Leftrightarrow$ $\mu<\infty$ and $|p|\rightarrow\infty$

\item [$B_{4}$] $\Leftrightarrow$ $\mu< \infty$ and $|\mu+3p|<\infty$
$\Leftrightarrow$ $\mu<\infty$ and $|p|<\infty$.
\end{description}
Of course we can translate these conditions to asymptotic behaviours
 in terms of $a,H$, depending on the value of $k$, for example,
\begin{enumerate}
\item If $k=0$, $\mu<\infty$ $\Rightarrow$ $H^{2}<\infty$, $a$ arbitrary

\item If $k=1$, $\mu<\infty$ $\Rightarrow$ $H^{2}<\infty$ and $a\neq 0$

\item If $k=-1$, $\mu<\infty$ $\Rightarrow$ $H^{2}-1/a^{2}<\infty$.
\end{enumerate}
As an example, we consider the sudden pressure singularity introduced by
Barrow
in \cite{ba04}.
This has a finite $a$ (condition $N_{2}$), finite $H$ (condition $S_{3}$),
finite $\mu$ but a divergent $p$ (condition $B_{3}$)
at $t_{s}$.

As another example, consider the  flat \textsc{FRW} model
containing dust and a scalar field studied in
\cite{mel}. The scale factor collapses at both
an initial (big bang) and a final time (big crunch). The Hubble
parameter and $\ddot{a}/a$  both blow up at the times of the big bang and
big crunch (cf. \cite{ck05}) leading to an $(S_{1},N_{1},B_{1})$ big bang
singularity and an $(S_{2},N_{1},B_{1})$ big crunch singularity, respectively.

\section{Generic results and examples}

It was recently shown by Ellis in \cite{ellis} that a \textsc{RW}
space with scale factor $a(t)$ admits a past closed trapped
surface if the following condition is satisfied: \be \label{cts}
\dot{a}(t)>\left|\frac{f'(r)}{f(r)}\right|, \ee with $f(r)=\sin
r,r,\sinh r$ for $k=1,0,-1$ respectively. Recall that a closed
trapped surface is a 2-surface with spherical topology such that
both families of incoming and outgoing null geodesics orthogonal
to the surface converge. Since our function $|D|$ can be written
in the form \be |D|=\frac{\sqrt{3}}{a^{2}(t)}
\left|\left(\dot{a}(t)-\frac{f'(r)}{f(r)}\right)
\left(\dot{a}(t)+\frac{f'(r)}{f(r)}\right)+
\frac{1}{f^{2}(r)}\right|, \ee we see that the condition for the
existence of a closed trapped surface becomes equivalent to the
following inequality: \be \label{d}
|D|>\frac{\sqrt{3}}{a^{2}(t)f^{2}(r)}. \ee We thus conclude that
collapse singularities (as predicted by the existence of a trapped
surface) are characterized by a divergent Bel-Robinson energy.

In this Section we provide necessary and sufficient conditions for
the occurrence of some of the triplets detailing the nature of the
singularities introduced above. These conditions are motivated
from studies of cosmological models described by exact solutions
in the recent literature. By exact solutions we mean those in
which all arbitrary constants have been given fixed values. We
expect the proofs of these results to be all quite
straightforward, for we have now \textit{already} identified  the
type of singularity that we are looking for in accordance with our
classification. Proving such results \emph{without} this knowledge
would have been a problem of quite a  different order.

The usefulness of the results proved below lies in that they
answer the question of whether the behaviours met in known models
described by exact solutions (which as a rule have a \textit{flat}
spatial metric ($k=0$)) continue to be valid in universes having
nonzero values of $k$ as well as described by solutions which are
more general than exact in the sense that some or all of the
arbitrary constants present remain arbitrary. The reason behind
this behaviour whenever it is met, lies in the fact that the
curvature term in the Friedman equation turns out to be usually
subdominant compared to the density term or in any case can not
alter the $H$ behaviour. For the purpose of organization we
present separately results about future and past singularities.

\subsection{Future singularities}
The first result of this subsection characterizes the future
singularity in phantom cosmologies \textit{irrespective} of the
value of the curvature $k$ and says that  the singularities in such
models can be milder than the standard big crunches and have necessarily
diverging pressure and the characteristic ``phantom" equation of state.
\begin{theorem} \label{2}
Necessary and sufficient conditions for an $(S_{2},N_{3},B_{1})$
singularity occurring at the finite future time $t_{s}$ in an
isotropic universe filled with a fluid with  equation of state
$p=w\mu$, are that $w<-1$ and $|p|\rightarrow\infty$ at $t_{s}$.
\end{theorem}

\begin{proof}
 Substituting the equation of state $p=w\mu$ in the
continuity equation $\dot{\mu}+3H(\mu+p)=0$, we have
\be
\label{phantom} \mu\propto a^{-3(w+1)}, \ee and so if $w<-1$ and
$p$ blows up at $t_{s}$, $a$ also blows up at $t_{s}$. Since
\be
H^{2}=\frac{\mu}{3}-\frac{k}{a^{2}},\quad
|D|^{2}=\frac{\mu^{2}}{3}, \quad
|E|^{2}=\frac{1}{12}\mu^{2}(1+3w)^{2}, \ee we conclude that at
$t_{s}$, $H$, $a$, $|D|$ and $|E|$ are divergent.

Conversely, assuming an $(S_{2}, N_{3}, B_{1})$
singularity at $t_{s}$ in an \textsc{FRW} universe with the equation of
state $p=w\mu$, we have from the $(B_{1})$ hypothesis that
$\mu\rightarrow\infty$ at $t_{s}$ and so $p$ also blows up at
$t_{s}$. Since $a$ is divergent as well, we see from
(\ref{phantom}) that $w<-1$.
\end{proof}

As an example, consider an exact solution which describes a flat,
isotropic phantom dark energy filled universe,  studied in
\cite{gonzales} given by
\be
\alpha={[{\alpha _{0}}^{3(1+w)/2}+\frac {3(1+w)\sqrt
{A}}{2}(t-t_{0})]}^{\frac {2}{3(1+w)}}, \ee where $A$ is a
constant. From (\ref{phantom}) we see that the scale factor, and
consequently $H$, blows up at the finite time
$$t_{s}=t_{0}+\frac{2}{3\sqrt {A}
(|w|-1)\alpha_{0}^{3(|w|-1)/2}}.$$ Then it follows that
$|E|^{2}=\frac{3}{4}H^{4}(1+3w)^{2}$ and $|D|^{2}=3H^{4}$ also
blow up at $t_{s}$. Therefore in this model the finite time
singularity is of type $(S_{2},N_{3},B_{1})$.

Next we focus on a generalization of a model, called graduated
inflation and originally  proposed by Barrow - see next
subsection, given in \cite{noj1}. Consider a flat \textsc{FRW}
filled with a fluid with equation of state $p+\mu=-B\mu^{\beta}$,
$\beta>1$. As $\mu\rightarrow\infty$, $$t\rightarrow
t_{0}+\frac{2}{\sqrt{3}\kappa B} \frac{\mu^{-\beta+1/2}}{1-2\beta}
\rightarrow t_{0},$$ where $\kappa^{2}=8\pi G$ and  $t_{0}$ is an
integration constant. It follows then that $|p|\rightarrow\infty$
at $t_{0}$. The scale factor is described by (\ref{scale1}) below
and it is therefore finite. However, $H$, $|D|$ and $|E|$ all
diverge at $t_{0}$ meaning that this is an $(S_{2},N_{2},B_{1})$
singularity in our scheme. As we shall show in the next
subsection, the introduction of a more general equation of state
has resulted in ``taming" the singularity of the graduated
inflationary type. This behavior is a special case of the
following general result which holds also in curved models.
\begin{theorem} \label{3}
A necessary and sufficient condition for an $(S_{2},N_{2},B_{1})$
singularity at $t_{s}$ in an isotropic universe filled with a
fluid with equation of state $p+\mu=-B\mu^{\beta}$, $\beta>1$, is
that $\mu\rightarrow\infty$ at $t_{s}$.
\end{theorem}
\begin{proof}
From  the continuity equation we have
\be
\label{scale1}
a=a_{0}\exp{\left(\frac{\mu^{1-\beta}}{3B(1-\beta)}\right)}, \ee
and so $a\rightarrow a_{0}$ as $t\rightarrow t_{s}$. Since
\be
H^{2}=\frac{\mu}{3}-\frac{k}{a^{2}},\quad
|E|^{2}=\frac{1}{12}(2\mu+3B\mu^{\beta})^{2},\quad
|D|^{2}=\frac{\mu^{2}}{3}, \ee we see that as $t\rightarrow
t_{s}$, $H$, $|E|$ and $|D|$ diverge, leading precisely to an
$(S_{2},N_{2},B_{1})$ singularity. The converse is obvious.
\end{proof}

Similarly one can prove:
\begin{theorem} \label{4}
A necessary and sufficient condition for an $(S_{3},N_{2},B_{3})$
singularity at $t_{s}$ in an isotropic universe filled with a
fluid with equation of state $p+\mu=-C(\mu_{0}-\mu)^{-\gamma}$,
$\gamma>0$, is that $\mu\rightarrow\mu_{0}$ at $t_{s}$.
\end{theorem}

\begin{proof}
Again using the continuity equation we find
\be
\label{scale2}
a\propto\exp\left\{{-\frac{{(\mu_{0}-\mu})^{\gamma+1}}{3C(\gamma+1)}}\right\},
\ee which is finite as $t\rightarrow t_{s}$. Also since
\begin{eqnarray}\label{electric}
H^{2}=\frac{\mu}{3}-\frac{k}{a^{2}}, \quad
|E|^{2}=\frac{1}{12}(2\mu+3C(\mu_{0}-\mu)^{-\gamma})^2, \quad
|D|^{2}=\frac{\mu^{2}}{3},
\end{eqnarray}
we see that as $t\rightarrow t_{s}$, $H$ and $|D|$ remain finite
whereas $|E|$ diverges, leading to an $(S_{3},N_{2},B_{3})$
singularity. The converse is immediate.
\end{proof}

The above equation of state is studied in \cite{noj1} for the case of a
flat universe. It follows then that as $\mu\rightarrow\mu_{0}$,
$t=t_{0}-\frac{(\mu_{0}-\mu)^{\gamma+1}}{\kappa C\sqrt{3\mu_{0}}(\gamma+1)}\rightarrow
t_{0}$ (an integration constant) and $|p|\rightarrow\infty$. The
scale factor is described by (\ref{scale2}) and it is therefore
finite at $t_{0}$ whereas $H<\infty$, $|D|<\infty$ and
$|E|\rightarrow\infty$ leading to an $(S_{3},N_{2},B_{3})$ singularity.
We therefore see that the nature of the singularity depends very
sensitively on even very mild changes in the equation of state.

A way to become more intimately acquainted with the nature of the
various singularities and to clearly distinguish their various
differences is to study the relative asymptotic behaviours of the
three functions that define the type of singularity. Using a
standard notation that expresses the behaviour of two functions
around the singularity at $t_{\ast}$, we can introduce a kind of
relative ``strength" in the singularity  classification. Let $f,g$
be two functions. We say that
\begin{enumerate}
\item $f(t)$ is \textit{much smaller } than $g(t)$, $f(t)<<g(t)$, if and only if \\
$\lim_{t\rightarrow t_{\ast}} f(t)/g(t)=0$

\item $f(t)$ is \textit{similar} to $g(t)$, $f(t) \sim  g(t)$, if and only if
$0<\lim_{t\rightarrow t_{\ast}} f(t)/g(t)<\infty$

\item $f(t)$ is \textit{asymptotic} to $g(t)$, $f(t)  \leftrightarrow g(t)$,
if and only if $\lim_{t\rightarrow t_{\ast}} f(t)/g(t)=1$,
\end{enumerate}

Using these notions
we find that standard radiation filled isotropic universes (with $k=0,\pm 1
$) have the asymptotic behaviours described by
$a<<H<<(|E|\leftrightarrow |D|)$, whereas the rest of the standard big
bang singularities have $a<<H<<(|E| \sim  |D|)$.

The phantom model of theorem (\ref{2}) has three possible
behaviours depending on the ranges of the $w$ parameter, namely,
if $-5/3<w<-4 /3$  we have $H<<a<<(|E| \sim |D|)$, if $-4/3<w<-1$
then $H<<(|E|\sim |D|)<<a$ whereas if $w<-5/3$ we have
$a<<H<<(|E|\sim |D|)$.

The fluid case of
theorem (\ref{3}) has the behaviour is $a<<H<<|D|<<|E|$.

The sudden singularity
of theorem (\ref{4}) is characterized by the behaviour
$(H \sim  |D| \sim  a)<<|E|$. If we identify a sudden singularity as the
kind of singularity occurring with a divergent $p$ while $a$ and $H$ remain
finite, then this asymptotic behaviour is the only possibility.

In the majority of the types of singularities that we have met in
our study the two quantities $|E|$, $|D|$  are the most wildly
diverging functions and therefore become the most dominant ones
asymptotically.

\subsection{Past singularities}
We now present two results on the nature of past time
singularities. These include the standard big bang ones but the
latter are obviously \emph{not} the only possibility.  The first
result given below predicts that the time singularity met in an
exact solution in the family of the so-called graduated
inflationary models first constructed in \cite{ba}, extends also
to open universes. The model consists of a flat isotropic
spacetime with a fluid with equation of state
$p+\mu=\gamma\mu^{3/4}$, $\gamma>0$, and admits the exact solution
\be
a=\exp\left(-\frac{16}{3^{3/2}\gamma^{2}t}\right),\quad\textrm{with}\quad
\mu=\frac{256}{9\gamma^{4}t^{4}}. \ee At $t=0$, $a\rightarrow 0$
($N_{1}$), $H\rightarrow\infty$ ($S_{1}$) and
$|E|^{2}=\frac{1}{12}(-2\mu+3\gamma\mu^{3/4})^{2}\rightarrow\infty$,
$|D|^{2}\rightarrow\infty$ $(B_{1})$. This behaviour can be made
to occur more generally due to the following result.

\begin{theorem} \label{5}
A necessary and sufficient condition for an $(S_{1},N_{1},B_{1})$
singularity at $t_{1}$ in an open or flat isotropic universe
filled with a fluid with equation of state
$p+\mu=\gamma\mu^{\lambda}$, $\gamma>0$ and $\lambda<1$, is that
$\mu\rightarrow\infty$ at $t_{1}$.
\end{theorem}
\begin{proof}
The continuity equation gives
\be
\label{scale}
a=a_{0}\exp{\left(\frac{\mu^{-\lambda+1}}{3\gamma(\lambda-1)}\right)},
\ee so that $a\rightarrow 0$ as $t\rightarrow t_{1}$. Since

\[ H^{2}=\frac{\mu}{3}-\frac{k}{a^{2}}=\frac{\mu}{3}-
k{a_{0}}^{-2}\exp{\left(\frac{-2\mu^{-\lambda+1}}{3\gamma(\lambda-1)}\right)}>0,
\]
\[ |E|^{2}=\frac{1}{12}(-2\mu+3\gamma\mu^{\lambda})^{2}, \]
\[|D|^{2}=\frac{\mu^{2}}{3},\]
we see that as $t\rightarrow t_{1}$ $H$, $|E|$ and $|D|$ diverge
provided that $k=0$ or $k=-1$. The converse is straightforward.
\end{proof}

The asymptotic strength of this singularity is
$a<<(|E|\leftrightarrow |D|)<<H$.

The second result that we prove now says that the strongest big
bang type singularities in universes with a massless scalar field
are produced due to the special form of the kinetic term. An exact
solution for the case of a flat spatial metric is derived in
\cite{fo} and is given by $$H=\frac{1}{3t}, \quad
\phi=\pm\sqrt{\frac{2}{3}}\ln\frac{t}{c}.$$ Since $a\propto
t^{1/3}$ we have that at $t=0$, $a\rightarrow 0$ ($N_{1}$),
$H=1/(3t)\rightarrow \infty$ ($S_{1}$),
$|E|^{2}=\dot{\phi}^{4}/3\rightarrow\infty$ and
$|D|^{2}=3H^{4}\rightarrow\infty$ ($B_{1}$). As it follows from
\cite{fo}, this exact solution represents the asymptotic behaviour
of a scalar field model if
$$\lim_{\phi\rightarrow\pm\infty}e^{-\sqrt{6}|\phi|}V(\phi)=0.$$

\begin{theorem} \label{6}
A necessary and sufficient condition for an $(S_{1},N_{1},B_{1})$
singularity at $t_{1}$ in an isotropic universe with a massless
scalar field is that $\dot{\phi}\rightarrow\infty$ at $t_{1}$.
\end{theorem}

\begin{proof}
From the continuity equation,
$
\ddot{\phi}+3H\dot{\phi}=0,
$
we have $\dot{\phi}\propto a^{-3}$, and if
$\dot{\phi}\rightarrow\infty$ then $a\rightarrow 0$. Since
\be
H^{2}=\frac{\mu}{3}-\frac{k}{a^{2}}\rightarrow\infty
%\frac{\dot{\phi}^{2}}{6}-k\dot{\phi}^{2/3},
\ee
$H$ becomes unbounded at $t_{1}$. In addition, since
\be
|D|^{2}=\frac{\mu^{2}}{3}=\frac{\dot{\phi}^{4}}{12}\rightarrow\infty,
\ee and
\be
|E|^{2}=\frac{1}{12}(\mu+3p)^{2}=\frac{\dot{\phi}^{4}}{3}\rightarrow\infty,
\ee at $t_{1}$, both $|D|$ and $|E|$ diverge there.

Conversely, assuming an $(S_{1},N_{1},B_{1})$ singularity at
$t_{1}$, we have (from $B_{1}$) that $\mu\rightarrow\infty$ and so
$\dot{\phi}^{2}\rightarrow\infty$ at $t_{1}$.
\end{proof}

The asymptotic strength of the singularity for the model described by both the exact solution
and the above theorem is $a<<H<<(|E| \sim  |D|)$.

A very mild type of singularity discovered recently in
\cite{noj1}. We shall see how our classification is able to pick
this up. The model is that of a flat \textsc{FRW} type filled with
a fluid with equation of state
$$p+\mu=-\frac{AB\mu^{2\beta-1}}{A\mu^{\beta-1}+B},\quad
0<\beta<1/2,$$ and for $\beta=1/5$ it admits the exact solution
$a=a_{0} e^{\tau^{8/3}}$. Then $H=(8/3)\tau^{5/3}$,
$\dot{H}=(40/9)\tau^{2/3}$ and $\ddot{H}=(80/27)\tau^{-1/3}$. Thus
as $\tau\rightarrow 0$, $a$, $\dot{a}$, $\ddot{a}$, $H$, $\dot{H}$
all remain finite whereas $\ddot{H}$ becomes divergent.

We can easily see that in this universe the Bel Robinson energy at
the initial time, $\mathcal{B}(0)$, is finite whereas \textit{its
time derivative} is
\be\dot{\mathcal{B}}(\tau)=3\left[2\frac{\ddot{a}}{a}(\ddot{H}+2H\dot{H})+
4\left(\frac{k}{a^{2}}+H^{2}\right) \left(-\frac{k
H}{a^{2}}+H\dot{H}\right)\right]\ee and thus
$\dot{\mathcal{B}}(\tau)\rightarrow\infty$ at $\tau\rightarrow 0$.
This is like a $B_4$ singularity in our scheme. Since the
derivative of the Bel-Robinson energy diverges, we may interpret
this singularity geometrically as one \emph{in the velocities of
the Bianchi (frame) field.} At $t_s$ the Bianchi field encounters
a cusp and its velocity diverges there. We believe this to be the
mildest type of singularity known to date.

\section{Discussion}
In this paper we have extended and refined the classification of the
cosmological singularities possible in an {\sc FRW} universe in general
relativity. A classification based entirely on the asymptotic behaviour of
the scale factor or the Hubble expansion rate cannot lead to complete
results and we have found it necessary to extend this scheme to include
the behaviour of the Bel-Robinson energy. In this case we have found that
the resulting behaviours of the three functions, $H$, $a$ and $\mathcal{B}$
taken together exhaust the types of singularities that are possible to
form during the evolution of an isotropic universe.

The resulting classification is described by triples of the form
$(S_{i},N_{j},B_{l})$ where the $S$ category monitors the asymptotic
behaviour of the expansion rate, closely related to the extrinsic
curvature of the spatial slices, $N$ that of the scale factor,
describing in a sense what the whole of space eventually does, while
$B$ describes how the matter fields contribute to the evolution of
the geometry on approach to the singularity. We know (cf.
\cite{chc02,ycb02}) that all these quantities need to be uniformly
bounded to produce geodesically complete universes. Otherwise, the
whole situation can be very complicated and we have exploited what
can happen in such a case when we consider a relatively simple
geometry as we have done in this paper.

Our scheme not only does cover all the recently discovered types of
singularities but it also predicts many possible new ones. For
example, in the case of a flat isotropic universe the classification
of \cite{noj1}   provides  us with four main types of singularity.
These four types can be identified with four particular
$(S_{i},N_{j},B_{k})$ triplets of our scheme. The `big rip' type
characterized by $a\rightarrow\infty$, $\mu\rightarrow\infty$ and
$|p|\rightarrow\infty$ at $t_{s}$ is an $(S_{2},N_{3},B_{1})$
singularity; the `sudden' singularity described by $a\rightarrow
a_{s}<\infty$, $\mu\rightarrow\mu_{s}<\infty$ and
$|p|\rightarrow\infty$ at $t_{s}$ is an $(S_{3},N_{2},B_{3})$;
further, the so-called type III singularity, namely $a\rightarrow
a_{s}<\infty$, $\mu\rightarrow\infty$ and $|p|\rightarrow\infty$ at
$t_{s}$ is clearly an $(S_{2},N_{2},B_{1})$ singularity, while, type
IV singularities with $a\rightarrow a_{s}<\infty$, $\mu\rightarrow
0$ and $|p|\rightarrow 0$ at $t_{s}$ belong to an
$(S_{3},N_{2},B_{4})$.

Having all possible singularity types in the form expounded in this
paper has the added advantage that we can consistently compare
between different types as we asymptotically approach the time
singularity. In this case the relative strength of the functions
describing the singularity type becomes an important factor to
distinguish between the possible behaviours. In this way we can have
a clear picture of how ``strong" a singularity can be when formed
during the evolution and also why such a type eventually arises in
any particular model. This charters the singularities in the
isotropic category, a necessary first step in an attempt to consider
the same classification problem in more complex situations.

It is natural to consider the extension of the work done in this
paper in the context of the more general anisotropic Bianchi models.
We believe that an analysis of this  more complicated case is still
feasible using the techniques of the present paper. In such contexts
the topology and spatial extent of the singular surfaces is expected
to play a role  even for the simplest non-trivial `vacua' such as
the Kasner or the Taub-NUT solutions and a complete elucidation of
these cases is a prerequisite to situations with matter fields. We
leave such matters to future work.

\section*{Acknowledgements}
We are very grateful to Y. Choquet-Bruhat for her many important
comments that have strongly influenced this work and to an anonymous
referee for kind suggestions. This work was supported by the joint
Greek Ministry of Education and European Union research grants
`Pythagoras' and `Heracleitus' and this support is gratefully
acknowledged.

\end{document}